\documentclass[12pt,aps,prb,preprint]{revtex4-1} % style for Physical Review B and AJP are similar

\usepackage{amsmath}    % need for subequations
\usepackage{graphicx}   % for figures
\begin{document}

\title{Asymptotic solution of a point-island model of irreversible aggregation with a time dependent deposition rate}
\author{Lionel Sittler}
\affiliation{Theoretische Physik, Fachbereich 8, Universit\"at Wuppertal \\and\\
 Universit\"at Duisburg-Essen, Germany}
 %\altaffiliation[Also at ]{home.}  %  optional
 %\affiliation{Clark University, Depar}
\date{\today}
\thanks{This work was partly supported by the DFG-grant WO 577/4-1.}
\begin{abstract}
In this paper we propose a solution for the time evolution 
of the
island density with irreversible aggregation and a time dependent
input of particle in the space dimensions $d=1,2$. For this
purpose we use the rate equation resulting from a generalized mean
field approach. 
%The most widely used technique 
A well-known technique
for growing
surfaces at the atomic scale is molecular beam epitaxy (MBE). Another
approach is the pulsed laser deposition method (PLD). The
main difference between MBE and PLD is that in the case
 of MBE we have a continuous rate of deposition $F$ of adatoms on the
 surface whereas in the case of PLD the adatoms are deposited during a pulse of a laser which is very
 short in comparison to the time span $T$ between
 the pulses.
The generalized mean field theory is a useful model for both MBE
and PLD  with the most simple approximation, point-like island.
 We show that the parameter $T$ distinguishes the MBE
regime from the PLD regime. We solve the rate equation for the PLD
regime.
 We consider the time evolution of the density of immobile islands. For large time $t\gg
 T$,
the PLD regime dominates the MBE regime and we find that the
density of immobile islands grows as $t^{1/2}$ whereas for MBE we
find the  known behavior of the density, $t^{1/3}$ for $d=2$
 and $t^{1/4}$ for $d=1$.
We illustrate this result with Monte-Carlo simulations for
$d=1,2$.
The author recognizes that in real experiments , some deviations from
 this  simple point island approximation in the rate equations could arise.
\end{abstract}
\maketitle
\section{Introduction}
Surface growth by pulsed deposition plays an important role in
the fabrication of thin films.
 In this technique the material is ablated by a
pulsed laser and then deposited in pulses so that bunches of many
atoms arrive at the surface simultaneously~\cite{PLD}. 

Pulsed laser deposition plays an important role in various applications
including the growth of ultra-hard carbon films, artificially
strained super-lattices, superconducting films, and multi-layered
complex structures~\cite{PLD,PLD1}. Alternatively, pulsed
deposition can be realized by chopping the flux of a continuous
source with a rotating shutter~\cite{chop}, independent of the experimental
process we call all these time-dependent depositions PLD.
% Compared to pulsed
%laser deposition, where the kinetic energy of the particles
%usually varies over a wide range, chopped deposition is
%particularly useful at low deposition energies of about $0.1
%\ldots 1$ eV~\cite{m}. 
Compared to ordinary MBE,
 those surfaces grown by PLD may exhibit a different surface
morphology~\cite{supra}. This observation triggered several
theoretical studies.
%For example, ~\cite{Narhe}
%investigated scaling properties of the island statistics in the
%c%coalescence regime of pulsed laser deposition at moderate and high
%deposition energies~\cite{Narhe}.
 Focusing on low energies there were numerical results
that PLD crosses over to MBE~\cite{MBEMF1, MBEMF2,MBEMF3,MBEMF4} and that the nucleation
density is characterized by unusual logarithmic scaling
laws~\cite{hinnemann}, which can be motivated in terms of local
scaling invariance with continuously varying
exponents~\cite{sittlerhinrichsen}. Moreover, it was shown that
the influence of a strong Ehrlich-Schwoebl barrier in PLD may lead
to a smoother surface compared to MBE~\cite{hinnemann}.
A  mechanism for film in PLD is described in ~\cite{aziz},
and review for more realistic models which connect with experiment 
~\cite{Mich,evans}. The mean-field approach is a convenient
approximation but its limitation is described in ~\cite{Blackman,Amar}
in $d=1$ and in ~\cite{evans,bartelt1} for $d=2$.
%, although this
%mechanism does not yet seem to explain the robustly improved
%quality of surfaces grown by pulsed laser deposition seen in
%recent experiments~\cite{Jenniches}.
So far theoretical studies of
pulsed deposition have been based mostly on numerical simulations.
The aim of the present work is to suggest a model
 for pulsed deposition which is expected to be valid in the
submonolayer regime, before coalescence, and in $d=1,2$ (i.e the
aggregation of two or more islands and an adatom is neglected,
only binary reactions are considered).
 To this aim we use a mean field approach for
PLD found in ~\cite{PLDHaye}  but we use the analytical rate of the
 generalized mean field in $d=1,2$~\cite{sittler} and we found the exact asymptotic
  solution of the time evolution of the total density of immobile islands.\\
%  MBE~\cite{MBEMF1,sittler} and replace the continuous flux $F$ by a periodically chopped flux $F(t)$.
Although the modification in the rate equation is rather small, we demonstrate that it
changes the entire scaling behavior already on the mean field
level.  The asymptotic solution for the immobile island density is
$t^{0.5}$ for $d=1,2$ whereas for MBE the solution is $t^{1/3}$
for $d=2$ and $t^{0.25}$ for $d=1$~\cite{wolf}. This confirms the
validity of the generalized Smoluchowski rate for $d=1$ not only
for systems without a source or with a constant source but also
for time-dependent input of particles~\cite{sittler}.
\section{Equations for PLD}
%------------------------------------------------------------------------------
%
We consider the following aggregation model. We assume:
\begin{center}
 \begin{enumerate}
 \item {\it Brownian motion} for monomers (with diffusion constant $D$).
 The islands of mass $k>1$ are immobile (i.e. $D_k=0$ for $k>1$).
\item    {\it Irreversible aggregation}, i.e. when an adatom $N_1$
contacts and thus aggregates irreversibly and eventually forms an
island with a larger mass $+1$.
 \end{enumerate}\end{center}
 
%For low spatial dimension ($d\leq 2$) the effect of the geometry
%of an island is at most logarithmic ~\cite{sittler}, therefore
% we assume that the islands
%are point-like in the Smoluchowski rates equation (the effective
%radius is $r_i=1$ with a convenient choice of unit of length).

In the first approximation we assume that the island are point-like
(e.g the effect of  their lateral dimension is negligible in comparison
to the effect of diffusion). This is a reasonable approximation up to 
coverage $Ft$ such that the ratio of the average island-size to the average 
island-distance  is much less than one.
Hence we assume that the effective
radius is $r_i=1$ with a convenient choice of unit of length.
Other parameters such as the capture number of incident
monomers $k_s$ can be also introduced  ~\cite{PLDHaye}.
The model like all the Smoluchowski models, 
are valid for small time.
The introduction of  other parameters ($r_i=i^{1/d_f}$,$k_s$..etc)
could improve the discrepancy we will find in the Monte-Carlo
simulations in $d=2$.
Of course for more larger time scale
such parameters would be relevant.
But for small time our model, with such approximation 
exhibits a scaling regime for PLD. \\
In our model we disregard the spatial density $n_k(r,t)$ of an island of mass $k$
for a given position $r$ and time $t$ and compute the spatial
average island density
$N_k=\frac{1}{d^{(d)}\mathbf{ r}} \int d^{(d)}\mathbf{r} n(\mathbf
{r},t) =<n(r,t)>_{r}$ . The time evolution of the average adatom
density $N_1$ and the immobile island density
$N=\sum_{k=2}^{+\infty}N_k$ are obtained by the generalized
Smoluchowski approach. The original model of Smoluchowski is valid
for $d\geq3$ and, with a logarithmic correction, for $d=2$. It is
possible to use the known Smoluchowski model for $d=1$, but the solutions
can exhibit big discrepancies. The approach found in
~\cite{sittler}  generalizes the Smoluchowski model with rates
,and the approach in ~\cite{bartelt2} generalizes for the size distributions.
These rates can be separated into two rates, the first is the
the mean field rate (found in the Smoluchowski model), and the second is called the
correlation rate. The time evolution in $d=2$ is (we consider the
reaction rate relevant for a small time scale, i.e. the mean-field
rate ~\cite{sittler}
 the similar equations were found in~\cite{wolf,PLD}):
\begin{eqnarray}
\label{MBE_d=2}
\dot{N}_1&=& -DN_1\bigl(2N_1+N\bigr) +F(t)\\
\nonumber
\dot{N} &=& DN_1^2,
\end{eqnarray}
and for a very anisotropic surface, when the diffusion of adatoms
is in one direction, we consider the $d=1$ case, hence the rate is
the correlation rate ~\cite{wolf,sittler}(the correlation rate is the rate
valid for more larger time than the mean-field rate, i.e. in $d=1$):
\begin{eqnarray}
\label{MBE_d=1}
\dot{N}_1 &=& -D(N+N_1)N_1\bigl(2N_1+N\bigr) +F(t)\\
\nonumber
\dot{N}&=& D(N+N_1)N_1^2.
\end{eqnarray}
We obtain the correlation rate by multiplying the mean-field rate
with the average radius $M_o=N+N_1$. The physical interpretation
of the average radius is the effective surface in interaction with
an arbitrary particle.
An identical approach for MBE in $d=1$
is described in~\cite{Blackman}
 which are in contradiction with ~\cite{Amar}, at least
 for the small time regime $N_1\ll N$.

We assume a time dependent deposition of particles $F(t)$ .
In ~\cite{flux,flux2,flux3} they consider a chopped deposition with 
two time scale, the time span of the deposition of adatoms
and the time between the pulses.
Considering that, contrary to the model proposed in ~\cite{flux}
 the time between two pulses is much larger than the time
between the depositions of adatoms during the pulses. In the limit
of very short pulses the flux is of the form:
\begin{equation}
\nonumber
F(t) = I\,\sum_{k=0}^{\infty}\delta(t-t_k) \,,
\end{equation}
where $I$ is the pulse intensity defined as the density of adatoms
deposited per pulse. The index $~k$ enumerates the pulses. For
simplicity we assume that the pulses are separated by a constant
time interval $T$, i.e.
\begin{equation}
\nonumber
t_k=kT.
\end{equation}
In order to find the density of adatoms $N_1$ and immobile islands
$N$ we approximate the equation first for small time $t$ and then
large time $t$. For small time $t$ there are few islands, i.e
$N_1\gg N$, the equation reads
\begin{eqnarray}
\label{PLDeqsmall}
\dot{N}_1 &=&I \sum_{k=0}^{\infty}\delta(t-Tk)
-2DN_1^{4-d}\\
\nonumber
 \dot{N}&=& DN_1^{4-d}.
\end{eqnarray}

After a few pulses there are more immobile islands than adatoms.
We perform the approximation $N_1 \ll N$ and the equation is:

\begin{eqnarray}
\label{PLDeqlarge}
\dot{N}_1&=&I \sum_{k=0}^{\infty}\delta(t-Tk)
-DN_1N^{3-d}\\
\nonumber
 \dot{N}&=& DN^2_1N^{2-d}.
\end{eqnarray}
In the following, we will solve theses pairs  of  equations
~(\ref{PLDeqsmall}) and ~(\ref{PLDeqlarge})  and compare them with
Monte-Carlo simulations. We notice that in MBE we have two
parameters $F,D$ and for PLD three parameters $T,I,D$ and two
unknown densities $N,N_1$.
 For MBE, after a rescaling we have a scale free
equation and for PLD the equations depend on a parameter which
distinguishes MBE from PLD.
The rate equation for  $d=2$ with the correlation rate is approximately~\cite{sittler}
 the rate equation for $d=1$:
\begin{eqnarray}
\label{PLDcorrelation1}
\nonumber
\dot{N}_1&=& -D(N+N_1)N_1\bigl(2N_1+N\bigr) +F(t)\\
\nonumber
\dot{N}&=& D(N+N_1)N_1^2,
\end{eqnarray}
 we will not
consider this case separately because the solutions are the same
as the case $d=1$.

\section{Solution of the rate equation}
The asymptotic solution for PLD is less simple than for MBE. In 
%Kevin
order to find the time evolution, let us
 consider the temporal evolution of the adatom density
between two pulses for $t \in (kT,(k+1)T)$.
$N_1$ is a quickly varying function in comparison to $N$. We then
assume that $N$ is constant between two pulses (i.e. for
$kT<t<(k+1)T$) and we perform an adiabatic
approximation~\cite{Arnold}. We solve the large time equation
~(\ref{PLDeqlarge}).
 For an arbitrary $N$ we have the formal solution
between the pulses(H.Hinrichsen proposed another method
for solving these equations and found the exponent $0.5$ for PLD):
\begin{equation}
\nonumber
N_1=A_k \exp(-D\int_{kT}^{t}N^{3-d}(u)du)=A_k \exp(-D(t-kT)N^{3-d}(t)),
\end{equation}
where $A_k$ is the amplitude between the pulses. Since each pulse
increases the adatom density by $I$, we have $\lim_{\epsilon
\rightarrow 0}(N_1((k+1)T+\epsilon)-N_1((k+1)T-\epsilon))=I,$
hence
 the amplitude $A_k$ follows the recurrence relation:
\begin{equation}
\noindent
A_{k+1}=A_k\exp(-DN_{kT}^{3-d}T)+I,
\end{equation}
where $N_{kT}=N(kT)$. The amplitude is:
\begin{equation}
\nonumber
A_k=I\frac{1-e^{-DN_{kT}^{3-d}kT}}{1-e^{-DN_{kT}^{3-d} T}}.
\end{equation}
The adatom density reads
\begin{equation}
\nonumber
N_1=I\frac{1-\exp(-(DN_{kT}^{3-d}kT))}{1-\exp(-DN_{kT}^{3-d}T)}
\exp(-DN^{3-d}(t-kT)).
\end{equation}
Let us now turn to the temporal evolution of the density
 $N(t)$ on time scales extending over many pulses. We
consider a time scale which is large in comparison with the time
$T$ between the pulses, hence $N$ is almost constant between the pulse.
 In order to compute the solution on such a
large scale we will use the adiabatic approximation:
\begin{eqnarray}
\label{algorithm}
\nonumber
\dot{N}(t)&=&\frac{N_{(k+1)T}-N_{kT}}{T}\\
\nonumber
&=&\frac{\int_{kT}^{(k+1)T}\dot{N}(u)du}{T}\\
\nonumber
&=&\frac{\int_{kT}^{(k+1)T}DN(u)^{2-d}N_1(u)^2du}{T}\\
\nonumber
&=&I^2\biggl(\frac{1-\exp(-DN_{kT}^{3-d}Tk)}{1-\exp(-DN_{kT}^{3-d}T)}\biggl)^2
\frac{(1-\exp(-2DN_{kT}^{3-d}T))}{2TN}.
\end{eqnarray}
%The equations wich satisfy $N$, for a fixed $N$,
 Two asymptotic behaviors can be observed: For a fixed $N$,
 the large time MBE, when  $Tk=t\rightarrow +\infty$
with $TN^{3-d}_{kT}\rightarrow 0$ we have
$\dot{N}=\frac{I^2}{DN^{4-d}T^2}$,
and the large time PLD, when $T\rightarrow +\infty$ we have $\dot{N}=\frac{I^2}{2TN}$.\\
\newline
\newline
 Then the asymptotic solution for the total immobile island density, for large time, reads
\begin{enumerate}
   \item {\it small $T$ and large $t$ the MBE regime for large time} $N=(\frac{(5-d)F^2t}{D})^{1/(5-d)}$
 %ref.~\cite{MBEMF1,MBEMF3,MBEMF4,sittler}
 \item {\it large $T$ and large $t$ the PLD regime} $N =(\frac{I^2t}{T})^{1/2}$.
 \end{enumerate}
For a given $T$ the time $t_c$ defined by the equation
 %Kevin
$DN(t_c)^{3-d}T\sim 1$, using one of the last equations for
$N$  then \begin{equation}
\label{critical}
 t_c\sim T^{\frac{1-d}{3-d}} 
\end{equation}
distinguishes the MBE regime from the PLD regime, i.e. for $t_c\gg 
T^{\frac{1-d}{3-d}}$ the PLD regime dominates. We notice that,
although all the equations depend on the dimension $d$, the PLD
asymptotic solution is independent of the dimension, i.e.
$t^{1/2}$ for $d=1,2$, and the PLD regimes in $d=1$ for an
arbitrarily small time dominates (we have no MBE regime),
whereas usually the solution is dependent on the dimension~\cite{sittler}.\\
We then compute the small time regime of Eq.~(\ref{PLDeqsmall}),
hence
\begin{eqnarray}
\label{sol}
\nonumber
\dot{N}_1 &=&-2DN_1^{4-d}\mbox{ for }kT<t<(k+1)T\\
\nonumber
\label{sol1} \dot{N}&=& DN_1^{4-d}.
\end{eqnarray}
The formal solution for $N_1$ is(we still assume the validity of the adiabatic
approximation e.g. the $N$ is constant between the pulses)
\begin{equation}
\nonumber
N_1=\big[\frac{5-d}{2D(t-kT)+B_k}\big]^{1/(5-d)},
\end{equation}
where $B_k$ is defined by 
$\lim_{\epsilon \rightarrow 0}(N_1((k+1)T+\epsilon)-N_1((k+1)T-\epsilon))=I$ then:
\begin{equation}
\nonumber
\big[\frac{5-d}{2DT+B_k}\big]^{1/(5-d)}=I+\big[\frac{5-d}{B_{k-1}}\big]^{1/(5-d)}.
\end{equation}
The asymptotic behavior for PLD is:
\begin{equation}
\nonumber
N_1\sim\frac{I}{\big[\frac{2D(t-kT)}{(5-d)I}+1\big]^{\frac{1}{5-d}}}\mbox{ for } T\gg 1
\end{equation}
and for MBE:
\begin{equation}
\nonumber
N_1\sim\frac{Ft}{\big[\frac{2D(t-kT)}{(5-d)Ft}+1\big]^{\frac{1}{5-d}}}\mbox{ for } T\ll 1,
\end{equation}
we have kept the first relevant correction term of the PLD and MBE
regimes.
 %Kevin
Two asymptotic behaviors can be observed for small time:
\begin{enumerate}
 \item {\it the small time PLD }
   $N\sim D\frac{I^{(4-d)}}{\big[\frac{2D(t-kT)}{(5-d)I}+1\big]^{(\frac{4-d}{5-d})}}t$
 \item {\it the small time MBE} when
$N\sim D\frac{F^{4-d}}{\big[\frac{2D(t-kT)}{Ft(5-d)}+1\big]^{(\frac{4-d}{5-d})}}\frac{t^{5-d}}{5-d}$.
 \end{enumerate}
For the integration of the equation for the density of immobile 
%Kevin
islands we consider a time scale larger than $T$, hence 
%Kevin
$\frac{1}{\frac{2D(t-kT)}{(5-d)I}+1}$ and
$\frac{1}{\frac{2D(t-kT)}{(5-d)Ft}+1}$ are almost constant.
We recover the small time MBE regime $N\sim \frac{F^{4-d}t^{5-d}}{5-d}$~\cite{wolf,MBEMF1}
\section{Numerical simulation}
\begin{figure}[h]
  \includegraphics[scale=0.8,angle=90]{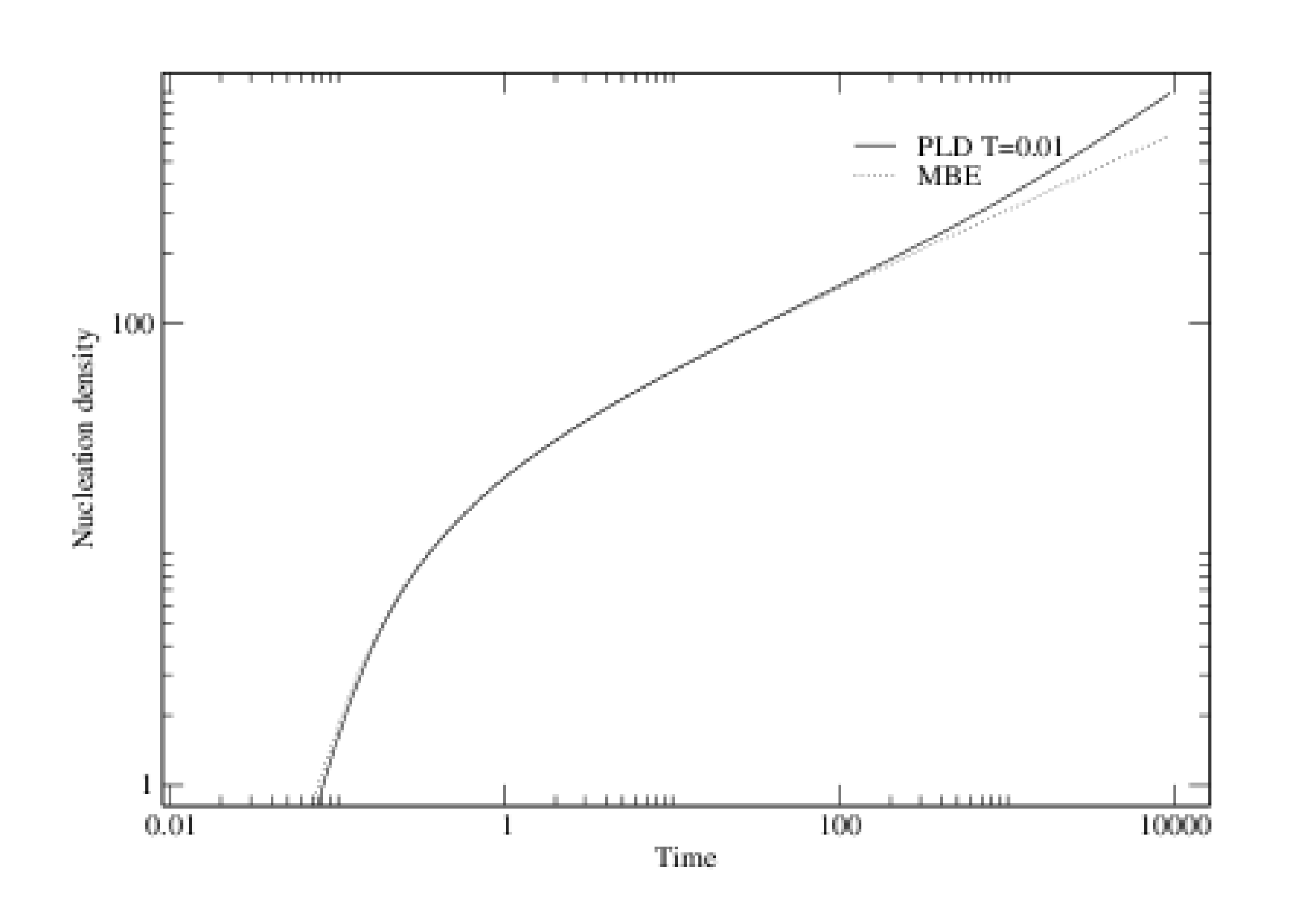}
 \caption
{\footnotesize  \label{a1} {\small   Numerical solution of the
Eq.~(\ref{MBE_d=2})with $d=2$ for PLD ($T=0.01$) and MBE (with the 
same parameters $F=I=D=1$, 
in arbitrary unit in order to illustrate the exponents).
 Notice the deviation of the PLD from MBE regime
at $t_c=T^{-1}=100$ see Eq.~\ref{critical}  }}
\end{figure}

\begin{figure}[h]
  \includegraphics[scale=0.9,angle=90]{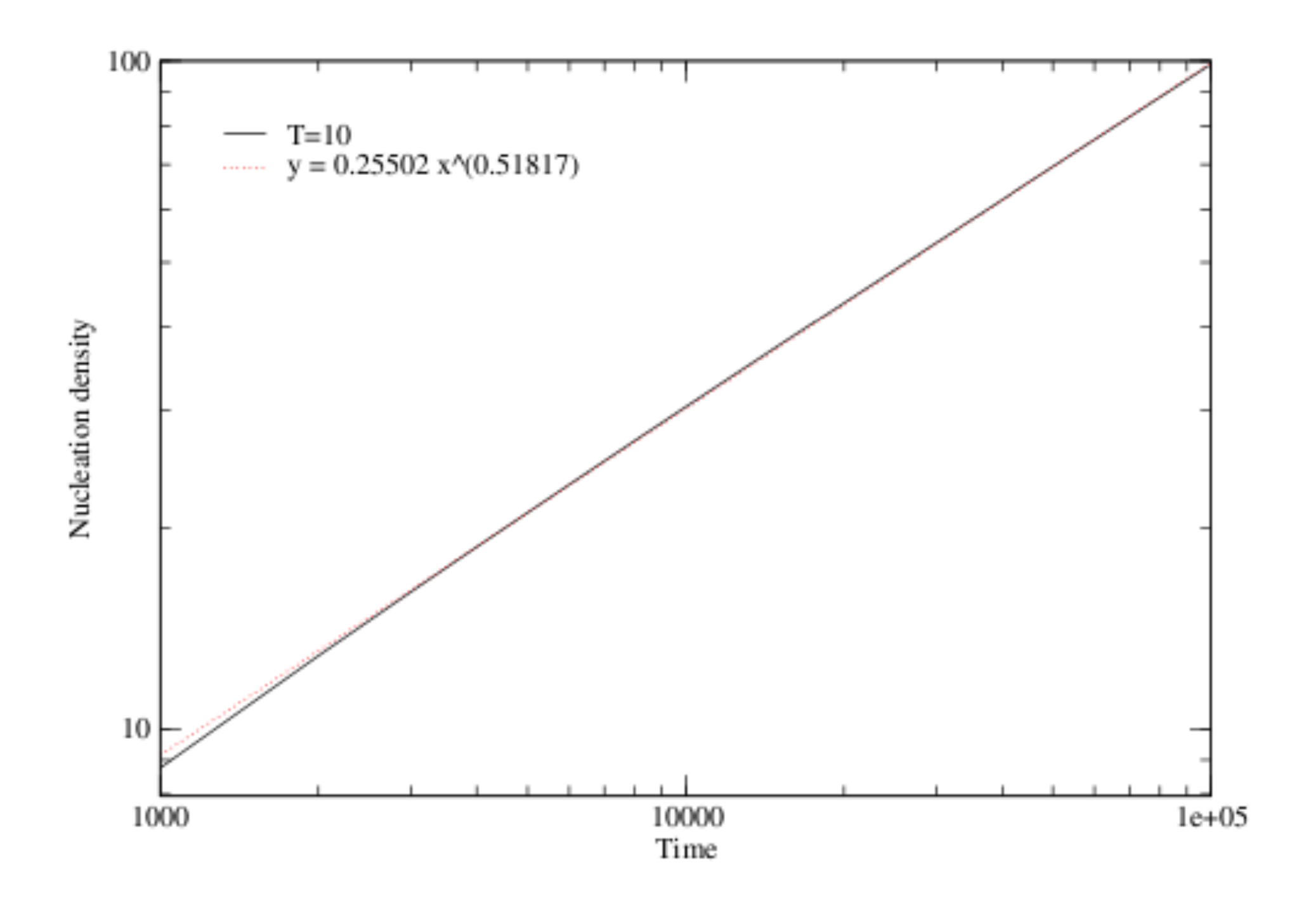}
 \caption
{\footnotesize
 \label{a2}{\small Kinetic Monte-carlo simulation in $d=1$. Nucleation density for $T=\infty$,
$d=1$. The best approximation
(the dashed line given by $Xmgrace$) has the exponent $\approx 0.518$}}.
\end{figure}

\begin{figure}[h]
  \includegraphics[scale=0.85,angle=90]{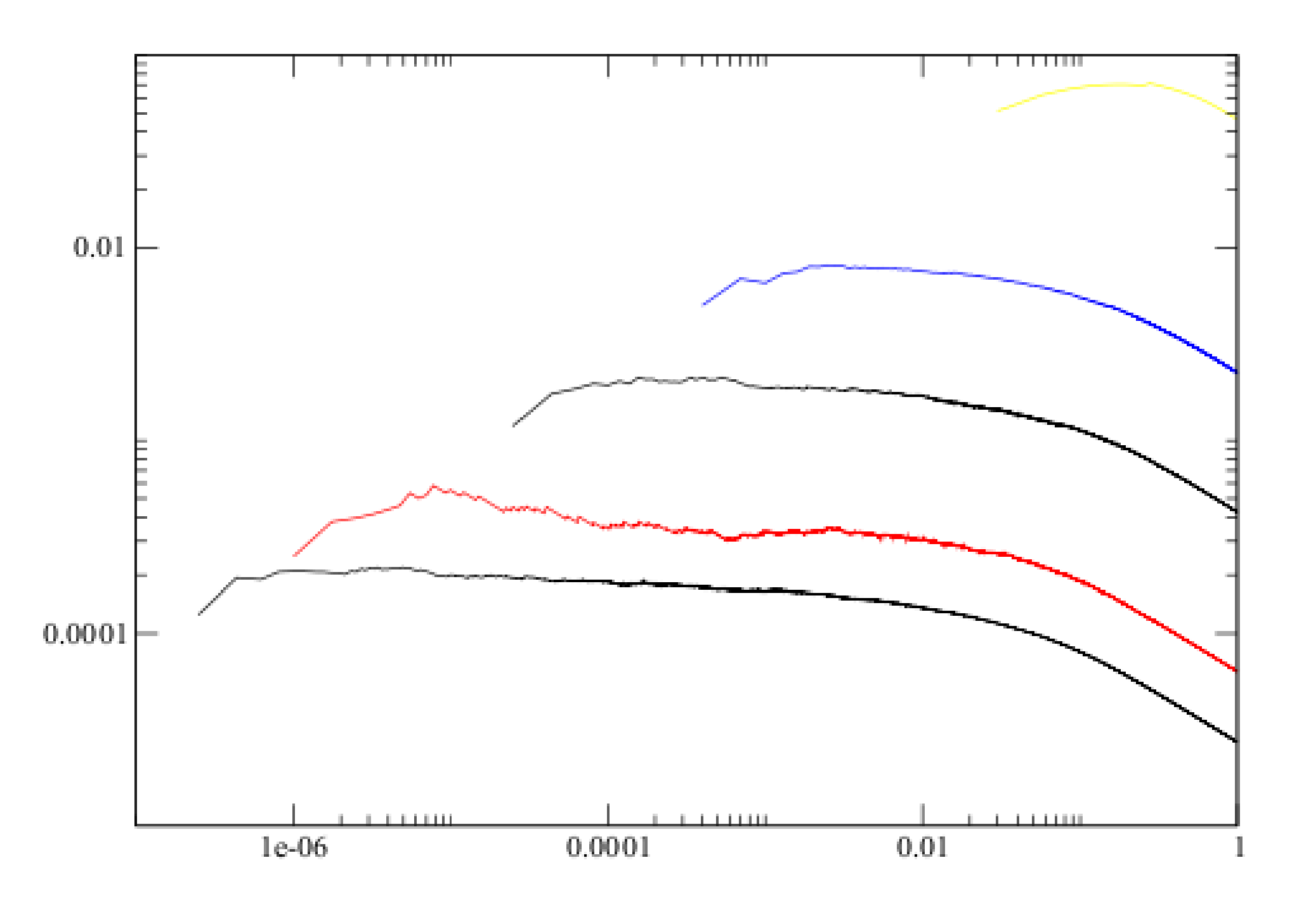}
 \caption
{\footnotesize
 \label{a3}{\small Kinetic Monte-carlo simulation in $d=2$. Nucleation density divided by
$t^{0.5}$ for $T=\infty$ and dimension
 $d=2$, with the density of deposited adatoms per pulse $I=10^{-7},10^{-6},10^{-5},10^{-4},10^{-2}$ from the lowermost to
the uppermost curve as function of time.}}
\end{figure}

 In the numerical simulation of the rates following equation Eq.~(\ref{PLDeqlarge})
 we show that the parameter $T$
controls the crossover between the MBE and PLD regimes (see Fig.~\ref{a1}).
We illustrate our approach by Kinetic Monte-Carlo simulations (KMC).
In ~\cite{Xu} a clear KMC realization has been presented,
 quite in accord with what has been presented in the present paper.
For the KMC simulations we have a d-dimensional lattice 
and process as following:
$I$ adatoms are deposited on each pulse, between 
the pulses we choose randomly an adatom
and moves it randomly in one of the  $2d$
possible directions.
 The comparison between the solution of the rate
equation with the Monte-Carlo simulation is performed with two
assumptions. First, we measure the nucleation density in the
Monte-Carlo simulation,
 i.e. the number of island creations
 per-unit surface(neglecting reaction rates with more than two
particles, the nucleation density is the same as $N$).
 Second, we consider the
  limit $T \rightarrow \infty$($t$ kept constant), and in order to simulate this limit we deposit $I$
 adatoms(only on free sites on the grid) and when all adatoms have merged
 (an adatom is the neighbor of an occupied site, sticks and does not diffuse
 anymore ) to an island or an adatom, i.e. $N_1(t)=0$
then $I$ adatoms are deposited (during the lapse time of
deposition of $I$ adatoms, the adatoms do not diffuse).
 In this regime the computation of the asymptotic value is simpler (see
   Fig.~\ref{a2} and Fig.~\ref{a3}(the evolution for $t>0.01$ cannot be describes by our model. The author supposes that it could be finite-size effect or coalescence)) and we get the non-trivial scaling behavior $t^{0.5}$
for the immobile island density $N$.
\section{Conclusion and outlooks}
In this paper we have shown that a pulsed input of particle yields
a very different solution of the immobile cluster density.
We propose a different type of crossover. We hope to find some experimental 
evidence of this scaling and crossover.
The author hopes that this approach will lead to similar result when for instance the 
deposition time of particle is not as small in comparison to the time span between 
two pulses~\cite{flux}. In this  case the rescaling of the rates equation will lead to a set
of equation with two parameters, the author expects to find that in this case the PLD regime 
dominates for large time. The PLD regime would be hence a regime which 
generalizes the MBE regime.
 \newpage
%
% ====================================================================


\begin{references}
% ====================================================================
%
\bibitem{PLD} D.B.Chrisey and G .K. Hubler(editors),Pulsed Laser Deposition, John wiley and Sons, New York(1994).
\bibitem{PLD1}R. G. Meyerand, Jr. and A. F. Haught, {\it Phys. Rev. Lett.}, {\bf 13}, 7Ð9 (1964).
\bibitem{chop} A. Tselev, A. Gorbunov, and W. Pompe,{\it Rev. Sci. Inst.}, 72, 2665-2672 (2001).% \bibitem{m} Michely T., private communication.
\bibitem{supra} T.Venkatesan, {\it Pulsed Laser Deposition : future Trends}
in D.B. Chrisey and G.K.Hubler(editors),{\it Pulsed Laser Deposition}, John wiley and Sons, New York(1994).
 \bibitem{MBEMF1}F.Westerhoff, L. Brendel
  and D.E. Wolf, in {\it Structure and Dynamics of heterogeneous Systems}, edited by P.Entel and D.E. Wolf ( World Scientific, singapore, 2000).
   \bibitem{MBEMF2} W. Matthew, {\it Epitaxial Growth}, (Academic, New York, 1975)
    \bibitem{MBEMF3}J.Y. Tao, {\it Materials Fundamentals of Molecular Beam Epitaxy}, (World Scientific, singapore, 1993).
    \bibitem{MBEMF4}J.G Amar, F. Family and P.-M. Lam, {\it Phys. Rev. B}{\bf 50}, 8781(1998).
 \bibitem{hinnemann}B. Hinnemann, H. Hinrichsen, and D. E. Wolf
 {\it Phys. Rev. Lett.} {\bf 87}, 135701 (2001). 
 \bibitem{sittlerhinrichsen}L.Sittler and H. Hinrichsen,{\it  J. Phys. A }{\bf35}, 10531-10538 (2002).
\bibitem{aziz} M.J. Aziz,{\it Appl. Phys. A}, {\bf 93},579 (2008). 
\bibitem{Mich}T.Michely and J.Krug , Atoms islands and mounds (Springer, 2004).
\bibitem{evans} J.W.Evans, R.Thiel, and M.C.Bartelt, {\it Surf. Sci. Rep.},{\bf 61} (2006).
 \bibitem{Blackman}J.A. Blackman, P.A. Mulheran, {\it Phys.Rev. B},{\bf 54}
 (1996) 11681.
\bibitem{Amar}J.G. Amar, M.N. Popescu, and F. Family, {\it Surf. Sci.}
,{\bf 491}, p.239-254 (2001). 
\bibitem{bartelt1} M.C.Bartelt and J.W.Evans, {\it Phys. Rev. B},{\bf 54} (1996) R17359.
 \bibitem{PLDHaye}A. C. Barato, H. Hinrichsen, and D. E. Wolf
{\it Phys. Rev. E}{\bf 77}, 041607 (2008).
\bibitem{sittler}L.Sittler {\it  J. Phys. A }{\bf 41}, 055005 (2008).
\bibitem{wolf} A. Pimpinelli, J. Vilain, and D.E. Wolf ,{\it Phys. Rev. Lett.}
, {\bf 69} 985, (1992).
\bibitem{bartelt2} M.C.Bartelt and J.W.Evans, {\it Phys. Rev.},{\bf B46} (1992) 12675.
 \bibitem{flux}P.Jensen and B. Niemeyer,{\it  Surf. Sci.}{\bf 384},L823-L827 (1997).
  \bibitem{flux2}
 S.Schinzer, M.Sokolowski,  M. Biehl and W.Kinzel{\it Phys. Rev. B}, {\bf 60}, 2893 (1999).
 \bibitem{flux3} N.Combes and P.Jensen{\it Phys. Rev. B},{\bf 57} 15553(1998).
\bibitem{Arnold} V.A. Arnol'd {\it Method of classical mechanics} Springer Verlag (1980).
%\bibitem{MBEMF3}J.Y. Tao, {\it Materials Fundamentals of Molecular Beam Epitaxy}, (World Scientific, singapore, 1993).
 %\bibitem{MBEMF4}L.H. Tang, {\it J. Phys. I} {\bf 3}, 935 (1993); J.G Amar, F. Family and P.-M. Lam, {\it Phys. Rev. B}{\bf 50}, 8781
 \bibitem{Xu}X.-J. Zhen, Bo Yan, Z. Zhu, B. Wu,and Y.-L. Mao.,{\it Thin Solid Films},{\bf 515} (2006) 2754Ð2759.
\end{references}
\end{document}